\documentclass{llncs}
\usepackage{times}
\usepackage{fullpage}

\usepackage{color,xcolor}
\usepackage{subfigure}
\usepackage{cleveref}
\usepackage{balance}
\usepackage{graphicx}
\usepackage{url}
\usepackage{multirow}
\usepackage{graphicx}
\usepackage{amssymb}
\usepackage{amsfonts}
\usepackage[small,bf]{caption}
\usepackage{wrapfig}
\usepackage{alltt}
\usepackage{booktabs}
\usepackage{xspace}
\usepackage{verbatim}
\usepackage{footnote}
\usepackage{epstopdf}
\definecolor{darkgreen}{RGB}{47,109,79}
\definecolor{darkblue}{RGB}{57,79,99}
\usepackage[bookmarks=false, colorlinks=true, plainpages = false,  linkcolor = darkgreen,   citecolor = darkgreen, urlcolor = darkblue, filecolor = blue]{hyperref}

\usepackage{flushend}
\newcommand{\precaption}{\vspace{0.1cm}}

\newcommand{\descr}[1]{\medskip\noindent\textbf{#1}}

\usepackage[hang,flushmargin,stable]{footmisc}

\usepackage{indentfirst}

\makeatletter
\def\url@foostyle{%
  \@ifundefined{selectfont}{\def\UrlFont{\rm}}{\def\UrlFont{\rmfamily}}}
\makeatother
\renewcommand{\subparagraph}{}
\usepackage[compact]{titlesec}
\titlespacing*{\section}{0pt}{*4}{5pt}
\titlespacing{\subsection}{0pt}{*4}{4pt}
\titlespacing{\subsubsection}{0pt}{*2.5}{0pt}

\begin{document}
\pagenumbering{arabic}

\title{Combating Fraud in Online Social Networks:\\ Detecting Stealthy Facebook Like Farms}

\author{Muhammad Ikram$^{\dag1}$, Lucky Onwuzurike$^{\dag2}$, Shehroze Farooqi$^{\dag3}$,\\[0.1ex]
Emiliano De Cristofaro$^2$, Arik Friedman$^1$, Guillaume Jourjon$^1$,\\[0.1ex]
Mohamed Ali Kaafar$^1$, M. Zubair Shafiq$^3$}
\institute{$^1$~Data61, CSIRO $\;\;^2$~University College London $\;\;^3$~University of Iowa}

\maketitle

\begin{abstract}

As businesses increasingly rely on social networking sites to engage with their customers, it is crucial to understand and counter reputation manipulation activities, including fraudulently boosting the number of Facebook page likes using {\em like farms}. To this end, several fraud detection algorithms have been proposed and some deployed by Facebook that use graph co-clustering to distinguish between genuine likes and those generated by farm-controlled profiles. However, as we show in this paper, these tools do not work well with stealthy farms whose users  spread likes over longer timespans and like popular pages, aiming to mimic regular users. 
We present an empirical analysis of the graph-based detection tools used by Facebook and highlight their shortcomings against more sophisticated farms. Next,  we focus on characterizing content generated by social networks accounts on their timelines, as an indicator of genuine versus fake social activity. We analyze a wide range of features extracted from timeline posts, which we group into two main classes: lexical and non-lexical. We postulate and verify that like farm accounts tend to often re-share content, use fewer words and poorer vocabulary, and more often generate duplicate comments and likes compared to normal users.
We extract relevant lexical and non-lexical features and and use them to build a classifier to detect like farms accounts, achieving significantly higher accuracy, namely, at least 99\% precision and 93\% recall.
\end{abstract}

\renewcommand{\thefootnote}{}
\footnotetext{\hspace*{-0.14cm}$^\dag$Authors contributed equally.}
\renewcommand{\thefootnote}{\arabic{footnote}}

\section{Introduction}
\label{sec: introduction}
Online social networks provide organizations and public figures with a range of tools to seamlessly reach out to, as well as broaden, their audience. Among these, {\em Facebook pages} make it easy to broadcast updates, publicize products and events, and get in touch with customers and fans. Facebook allows page owners to promote their pages via targeted advertisement. This constitutes one of the primary sources of revenue for Facebook, as its advertising platform is reportedly used by 2 million small businesses, out of the 40 million which have active pages~\cite{facebookpagecount}.

At the same time, as the number of likes on a Facebook page is considered a measure of its popularity~\cite{carter13like}, an ecosystem of so-called {\em ``like farms''} has emerged that offer paid services to artificially inflate the number of likes on Facebook pages. These farms often rely on networks of fake and compromised accounts, as well as incentivized collusion networks where users are paid for actions from their account~\cite{viswanath14tanomaloussocialnetwork}.
In prior work~\cite{decristofaro14facebooklikefarms}, we showed that some like farms follow a na\"ive approach with a large number of accounts liking target pages within a short timespan. Whereas, others exhibit a {\em stealthier} behavior, gradually spreading likes over longer timespans, aiming to evade fraud detection algorithms. We found that only a handful of like farm accounts were detected by Facebook \cite{decristofaro14facebooklikefarms}.

Facebook discourages users to buy fake likes, warning that they {\em ``can be harmful to your page''}\footnote{See \url{https://www.facebook.com/help/241847306001585}.} and routinely launches clean-up campaigns to remove fake accounts, including those engaged in like farms. Aiming to counter like farms, researchers as well as Facebook have been working on tools to detect fake likes (see Section~\ref{sec: related work}). One currently deployed tool is CopyCatch, which detects lockstep page like patterns by analyzing the social graph between users and pages, and the times at which the edges in the graph are created~\cite{beutel2013copycatch}. Another one, SynchroTrap, relies on the fact that malicious accounts usually perform loosely synchronized actions in a variety of social network context, and can cluster malicious accounts that act similarly at around the same time for a sustained period of time~\cite{cao14synchrotrap}. The issue with these methods, however, is that stealthier (and more expensive) like farms can successfully circumvent them by spreading likes over longer timespans and liking popular pages to mimic normal users.

As a consequence, in this paper, we set to characterize the liking patterns of accounts associated with like farms and systematically evaluate the effectiveness of graph co-clustering fraud detection algorithms~\cite{beutel2013copycatch,cao14synchrotrap} in correctly identifying like farm accounts. We show that these tools incur  significantly high false positives rates for stealthy farms, as their accounts mimic normal users. Next, we investigate the use of timeline information, including lexical and non-lexical characteristics of user posts, to improve the detection of like farm accounts. We crawl and analyze timelines of user accounts associated with like farms as well as a baseline of normal user accounts. Our analysis of timeline information highlights several differences in both lexical and non-lexical features of baseline and like farm users. In particular, we find that timeline posts by like farm accounts have 43\% fewer words, a more limited vocabulary, and lower readability than normal users' posts. Moreover, like farm posts generate significantly more comments and likes, and a large fraction of their posts consists of non-original and often redundant ``shared activity'' (i.e., repeatedly sharing posts from other users, articles, videos, and external URLs).

Based on these timeline-based features, we train three classifiers using supervised two-class support vector machines (SVM)~\cite{Muller01anintroduction} and evaluate them using our ground-truth dataset.  Our first and second classifiers use, respectively, lexical and non-lexical features extracted from timeline posts, while the third one uses both. Our evaluation shows that the latter can accurately detect like farms accounts, achieving up to 99--100\% precision and 93--97\% recall. Finally, we generalize our approach using other state-of-the-art classifier algorithms, namely, decision tree~\cite{dtree}, AdaBoost~\cite{adaboost}, kNN~\cite{knn}, random forest~\cite{Breiman:rf}, and na\"ive Bayes~\cite{zhang2004optimality}, and empirically confirm that the SVM classifier achieves higher accuracy across the board.

\descr{Paper Organization.}
The rest of the paper is organized as follows. Next section introduces the datasets used in our experiments, while
Section~\ref{sec:coclustering} evaluates the accuracy of state-of-the-art co-clustering techniques to detect like farm accounts in our datasets. Next, we study timeline based features (both non-lexical and lexical) in Section~\ref{sec:characterizing}, and evaluate classifiers built using these features in Section~\ref{sec:detection}. After reviewing related work in Section~\ref{sec: related work}, the paper concludes in Section~\ref{sec:conclusion}.

\section{Data}
\label{sec: data}

\descr{Previous Campaigns.} Our starting point are the Facebook accounts gathered as part of our prior work \cite{decristofaro14facebooklikefarms},
which presented an exploratory analysis of Facebook like farms using honeypots. Specifically, we created 13 Facebook pages called {\em ``Virtual Electricity"} and, while keeping them empty (i.e., no posts/pictures), promoted eight of them using popular like farms and five using Facebook ``page like'' ads. The eight like farm campaigns employed BoostLikes.com (BL), SocialFormula.com (SF), AuthenticLikes.com (AL), and MammothSocials.com (MS), each with one campaign targeting worldwide users and one targeting users in the USA, while the five Facebook campaigns respectively targeted users in the USA, France, India, Egypt, and worldwide. In the rest of the paper, we use the campaign acronyms followed by the target audience, e.g., SF-ALL denotes the SocialFormula.com campaign targeting worldwide users. Note that BL-ALL and MS-ALL did not actually deliver any likes, even though they were paid.

Overall, our campaigns \cite{decristofaro14facebooklikefarms} garnered 5,918 likes from 5,616 unique users: 1,437 unique accounts from Facebook ad campaigns and 4,179 unique accounts from the like farm campaigns. Note that some users liked more than one honeypot pages. After a few months, we checked how many accounts had been closed or terminated and found that 624/5,616 accounts (11\%) were no longer active.

\descr{New Data Collection.} In August 2015, we began to crawl the pages liked by each of the 4,179 like farm users from~\cite{decristofaro14facebooklikefarms}, using the Selenium web driver.\footnote{\url{http://docs.seleniumhq.org/projects/webdriver/}} We also collected basic information associated with each page, such as the total number of likes, category, and location, using the page identifier. Unlike in our previous study, we now also crawled the timelines of the like farm accounts. Specifically, we collected timeline posts (up to a maximum of 500 recent posts), the comments on each post, as well as the associated number of likes and comments on each post.

Besides some accounts having become inactive (376), we also could not crawl the timeline of 24 users who had restricted the visibility of their timeline. Moreover, Facebook blocked all the accounts we used for crawling, so we stopped our data collection before we could completely finish our data collection, hence, we missed further 109 users. In summary, our new data consists of 3,670 users out of the initial 4,179 users (88\%) gathered in~\cite{decristofaro14facebooklikefarms}. We collected more than 234K posts (messages, shared content, check-ins, etc) for these accounts, noticing that 72\% of them had at least 10 publicly visible posts.

We also rely on a sample of 1,408 random accounts previously gathered by Chen et al.~\cite{Chen2012}, which we use to form a baseline of ``normal'' accounts. For each of these accounts, we again collected posts from their timeline, their page likes, and information from these pages. 53\% of the accounts had at least 10 visible posts on the timeline, and in total we collected about 35K posts.

\begin{table}[t]
\small
\centering
\tabcolsep=0.125cm
\begin{tabular}{lrrrr}
\toprule
 & & {\bf \#Pages} &  {\bf \#Pages Liked} &   \\[-0.5ex]
{\bf Campaign} & {\bf \#Users}  & {\bf Liked} & {\bf (Unique)} & {\bf \#Posts} \\
\midrule
BL-USA & 583 & 79,025 & 37,283 & 44,566\\ %
SF-ALL & 870 & 879,369 & 108,020 &46,394 \\ %
SF-USA  & 653 & 340,964 & 75,404 & 38,999 \\ %
AL-ALL  & 707 & 162,686 & 46,230 &61,575\\  %
AL-USA   & 827 & 441,187 & 141,214 & 30,715\\ %
MS-USA  & 259 & 412,258 & 141,262 & 12,280\\  %
{\em Tot. Farms} & {\em 3,899} & {\em 2,315,489} & {\em 549,413}  & {\em 234,529}\\
\midrule
{Baseline~\cite{Chen2012}}  & 1,408 & 79,247 & 57,384 & 34,903\\
\bottomrule
\end{tabular}
\precaption
\caption{Overview of the datasets used in our study.}
\label{tbl:measurements}
\end{table}

Table~\ref{tbl:measurements} provides a summary of the data used in this paper. Note that users who like more than one of the honeypot Facebook pages are included in all rows, hence the disparity between 3,670 (unique users) and 3,899. Overall, we gathered information from about 600K unique pages, liked by 3,670 like farm accounts and 1,408 normal accounts, and around 270K timeline posts.

\descr{Ethical Considerations.} Note that we collected openly available data such as (public) profile and timeline information, as well as page likes. Also, all data was encrypted at rest and has not been re-distributed. No personal information was extracted as we only analyzed aggregated statistics. We also consulted our Institutional Review Board (IRB), which  classified our research as exempt.

\section{Limitations of Graph Co-Clustering Techniques}
\label{sec:coclustering}
Aiming to counter fraudulent activities, including like farms, Facebook has recently deployed detection tools such as CopyCatch~\cite{beutel2013copycatch} and SynchroTrap~\cite{cao14synchrotrap}. These tools use graph co-clustering algorithms to detect large groups of malicious accounts that like similar pages around the same time frame. However, as highlighted in our prior work~\cite{decristofaro14facebooklikefarms}, some stealthy like farms deliberately modify their behavior in order to avoid synchronized patterns, which might reduce the effectiveness of these detection tools. Specifically, while several farms use a large number of accounts (possibly fake or compromised) liking target pages within a short timespan, some spread likes over longer timespans and onto popular pages aiming to circumvent fraud detection algorithms.

\descr{Experiments.} We now set to evaluate the effectiveness of user-page graph co-clustering algorithms. We use the labeled dataset of 3,670 users from six different like farms and the 1,408 baseline users, and employ a graph co-clustering algorithm to divide the user-page bipartite graph into distinct clusters~\cite{Kluger03biclustering}. Similar to CopyCatch~\cite{beutel2013copycatch} and SynchroTrap~\cite{cao14synchrotrap}, the clusters identified in the user-page bipartite graph represent near-bipartite cores, and the set of users in a near-bipartite core like the same set of pages. Since we are interested in distinguishing between two classes of users (like farm users and normal users), we set the target number of clusters at 2.

\begin{table}[t]
\small
  \begin{center}
\tabcolsep=0.125cm
    \begin{tabular}{lrrrrrrr}
      \toprule
      {\bf Campaign} & {\bf TP} & {\bf FP} & {\bf TN} & {\bf FN} & {\bf Precision} & {\bf Recall} & {\bf F1-score} \\
      \midrule
AL-USA & 681 & 9 & 569 & 4 & 98\% & 99\% & 99\% \\
AL-ALL & 448 & 53 & 527 & 1 & 89\% & 99\% & 94\% \\
{\bf BL-USA} & 523 & 588 & 18 & 0 & \textbf{47\%} & 100\% & {\bf 64\%} \\
SF-USA & 428 & 67 & 512 & 1 & 86\% & 100\% & 94\% \\
SF-ALL & 431 & 48 & 530 & 2 & 90\% & 99\% & 95\% \\
MS-USA & 201 & 22 & 549 & 2 & 90\% & 99\% & 93\% \\
      \bottomrule
    \end{tabular}
\precaption
      \caption{Effectiveness of the graph co-clustering algorithm.}
          \label{tab: clustering accuracy}
  \end{center}
\end{table}

\begin{figure*}[!h]
\centering
\subfigure[AL-USA]
{\includegraphics[width=.35\columnwidth]{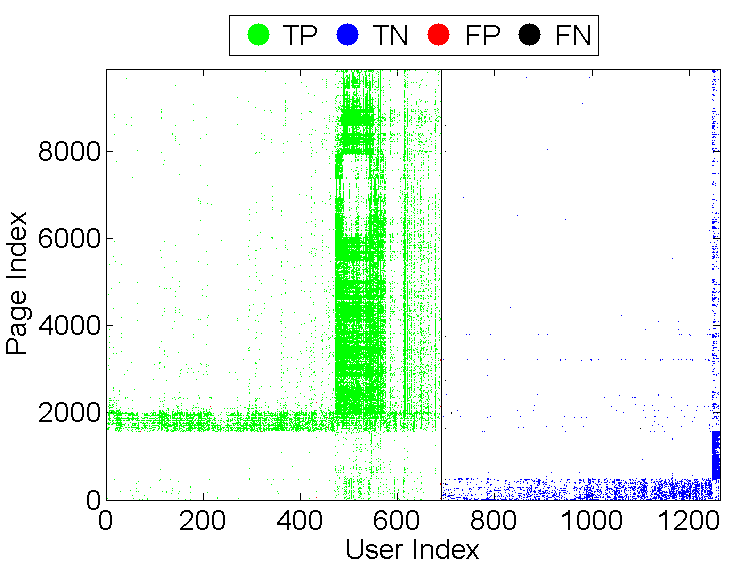}}
~
\subfigure[AL-ALL]
{\includegraphics[width=.35\columnwidth]{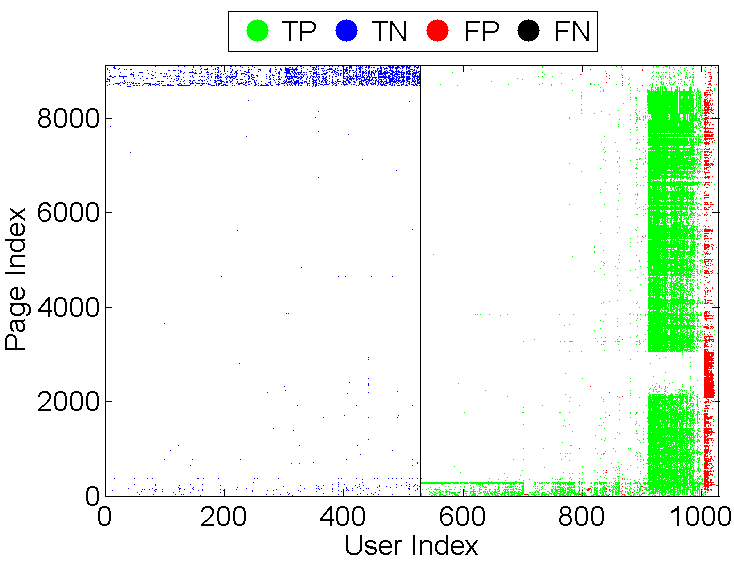}}\\
\subfigure[BL-USA]
{\includegraphics[width=.35\columnwidth]{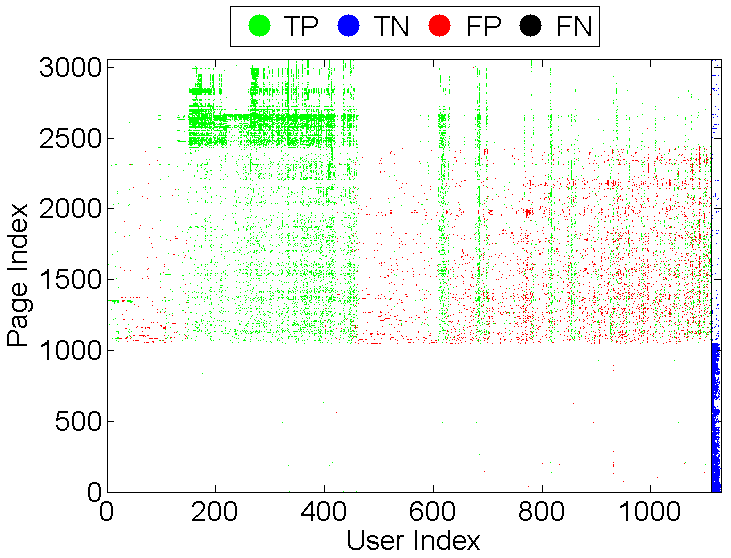}}
~
\subfigure[SF-USA]
{\includegraphics[width=.35\columnwidth,height=4.65cm]{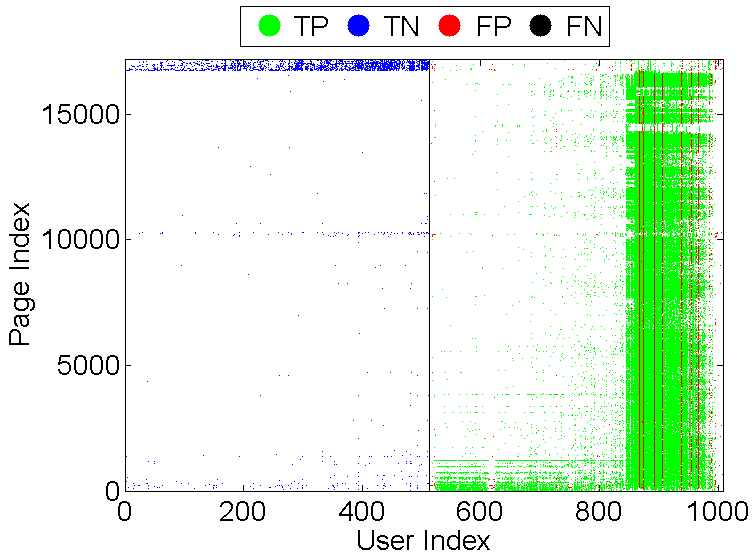}}\\
\subfigure[SF-ALL]
{\includegraphics[width=.35\columnwidth]{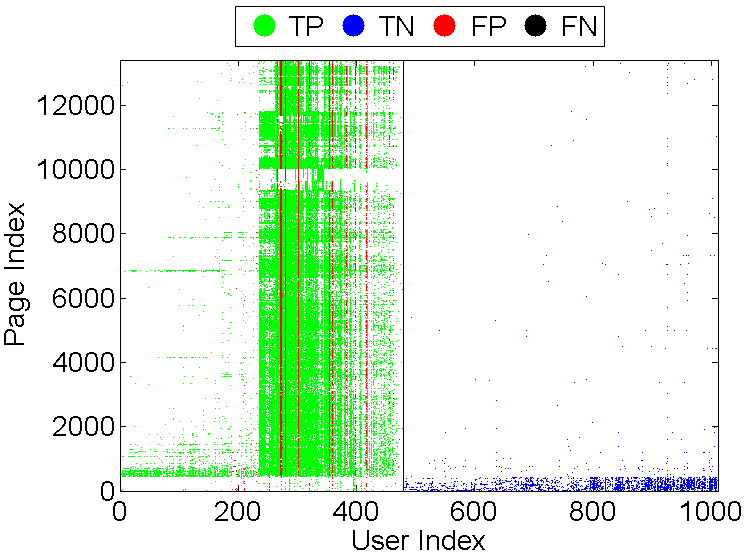}}
~
\subfigure[MS-USA]
{\includegraphics[width=.35\columnwidth]{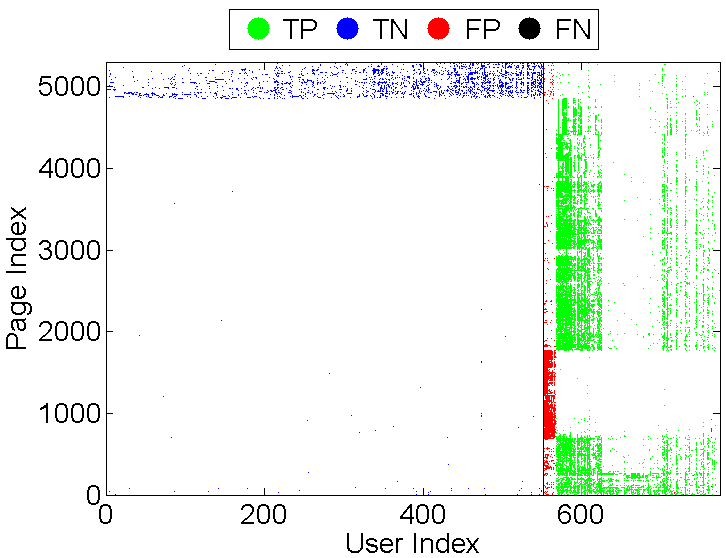}}
\vspace{-0.2cm}
\caption{Visualization of graph co-clustering results. The vertical black line indicates the separation between two clusters. We note that the clustering algorithm fails to achieve good separation leading to a large number of false positives (red dots).}
\label{fig: clustering results}
\vspace{0.4cm}
\end{figure*}

\descr{Results.} In Table~\ref{tab: clustering accuracy}, we report the ROC statistics of the graph co-clustering algorithm -- specifically, true positives (TP), false positives (FP), true negatives (TN), false negatives (FN), Precision: $(TP)/(TP+FP)$, Recall: $(TP)/(TP+TN)$, and F1-score, i.e., the harmonic average of precision and recall. Figure~\ref{fig: clustering results} visualizes the clustering results as user-page scatter plots. The x-axis represents the user index and the y-axis the page index.\footnote{To ease presentation, we exclude users and pages with less than 10 likes.} The vertical black line marks the separation between two clusters. The points in the scatter plot are colored to indicate true positives (green), true negatives (blue), false positives (red), and false negatives (black).

\descr{Analysis.} We observe two distinct behaviors in the scatter plots: (1) {\em ``liking everything''} (vertical streaks), and (2) {\em ``everyone liking a particular page''} (horizontal streaks). Both like farm and normal users exhibit vertical and horizontal streaks in the scatter plots. 

While the graph co-clustering algorithm neatly separates users for AL-USA, it incurs false positives for other like farms. In particular, the co-clustering algorithm fails to achieve a good separation for BL-USA, where it incurs a large number of false positives, resulting in 47\% precision. Further analysis reveals that the horizontal false positive streaks in BL-USA include popular pages, such as ``Fast \& Furious" and ``SpongeBob SquarePants,'' each with millions of likes. We deduce that stealthy like farms, such as BL-USA, use the tactic of liking popular pages aiming to mimic normal users, which reduces the accuracy of the graph co-clustering algorithm.

Our results highlight the limitations of prior graph co-clustering algorithms in detecting fake likes by like farm accounts. We argue that fake liking activity is challenging to detect when only relying on monitoring the liking activity due to the increased sophistication of stealthier like farms. Therefore, as we discuss next, we plan to leverage the characteristics of timeline features to improve accuracy.

\section{Characterizing Timeline Features}
\label{sec:characterizing}
We now set to design and evaluate timeline-based detection of like farm accounts. We start by characterizing timeline activities with respect to two categories of features, {\em non-lexical} and {\em lexical}. We do so aiming to identify the most distinguishing features to be used by machine learning algorithms in order for accurately classifying like farms accounts and normal user accounts.

\begin{figure}[h]
\centering
\vspace{-0.25cm}
\includegraphics[width=0.45\columnwidth]{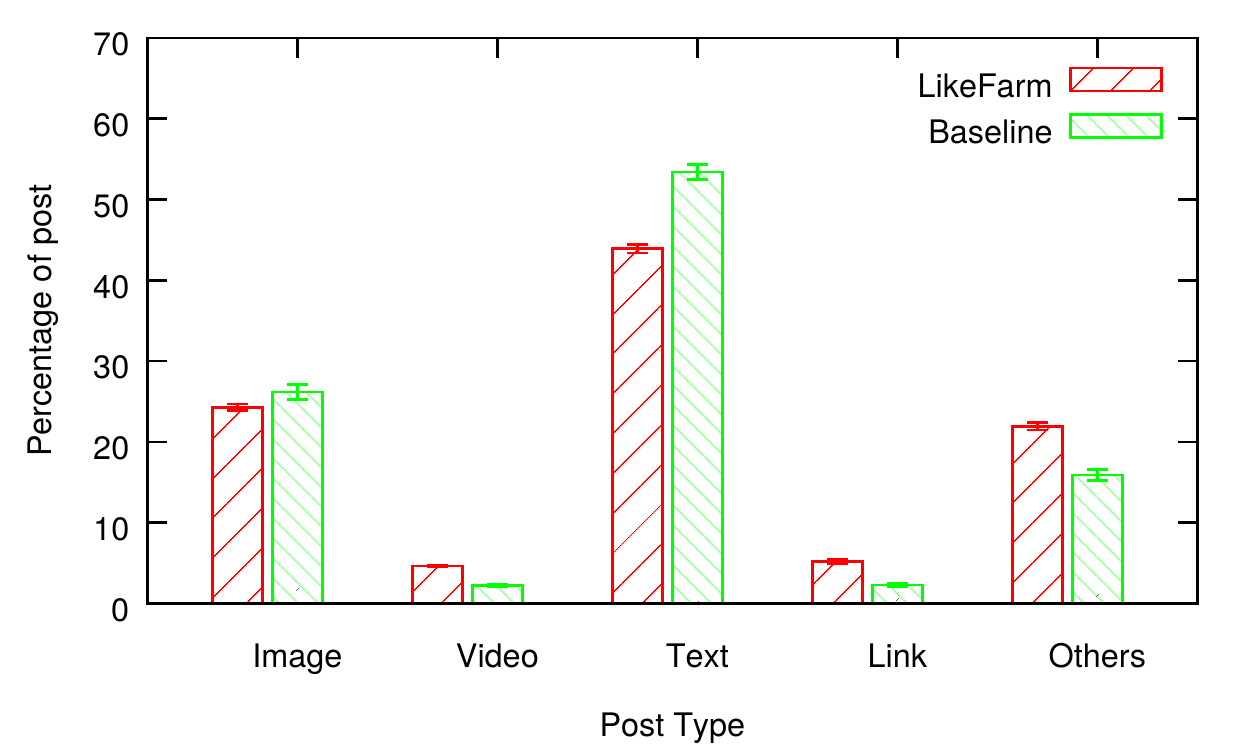}
\vspace{-0.25cm}
\caption{Distribution of types of posts.}
\vspace{-0.6cm}
\label{fig:posttype}
\end{figure}

\descr{Types of Posts.} 
We start by analyzing how users interact with the posts on their timeline. 
Figure~\ref{fig:posttype} plots the distribution of {\em types} of posts on users' timelines. More than 50\% of those made by baseline users are text, whereas, for like farm users this ratio is less than 44\% as they post more web links and videos. Note that ``Others'' include shared posts, Facebook actions such as `listening to', `traveling to', `feeling', and life events like `in a relationship', and `married'. We find that this category includes about 22\% of posts for like farm users and about 16\% of posts for baseline users.

\begin{figure*}[t]
	\centering
\subfigure[]{\includegraphics[width=0.35\columnwidth]{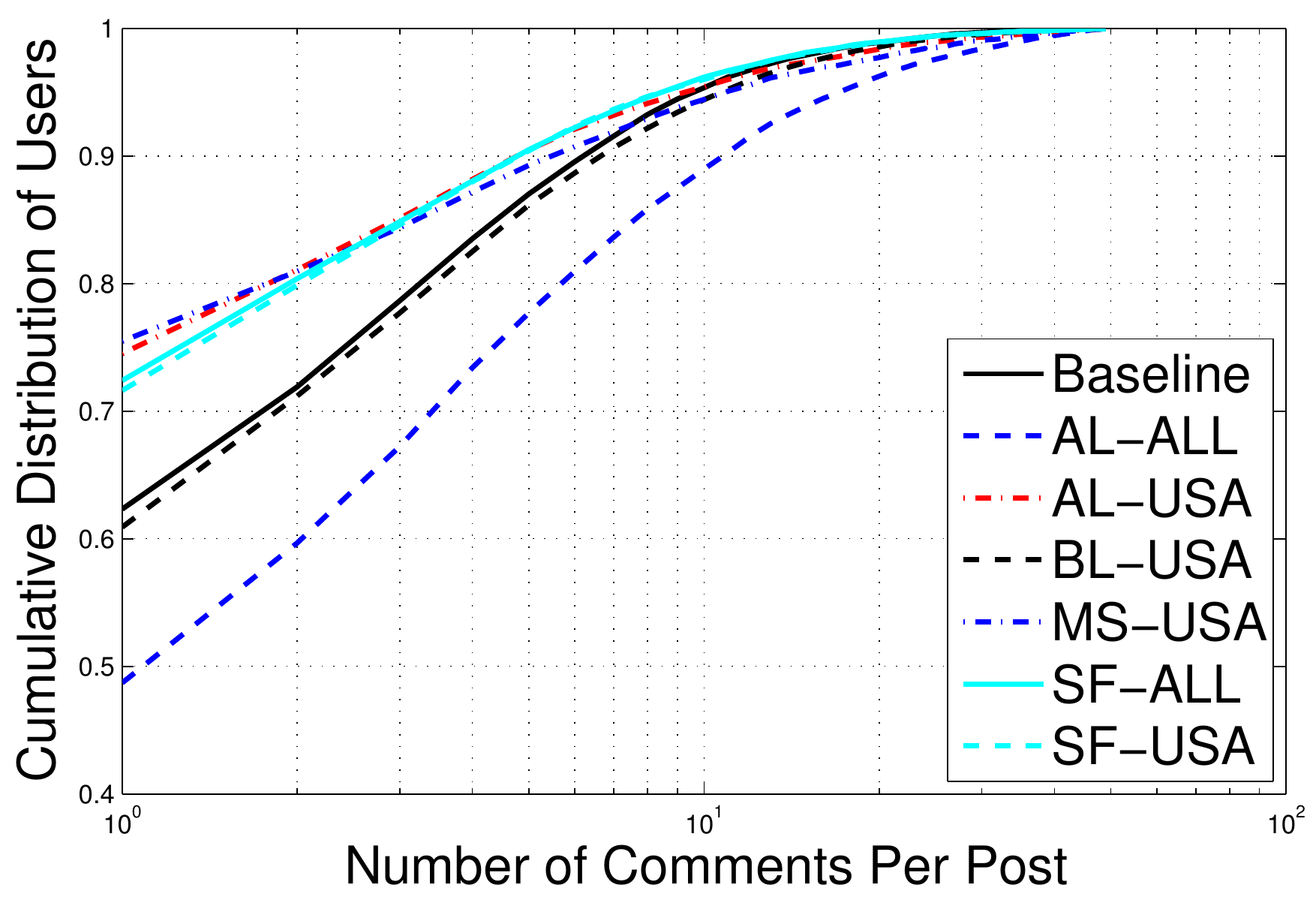}\label{fig:comperpost}}
~\hspace{-0.2cm}
\subfigure[]{\includegraphics[width=0.35\columnwidth]{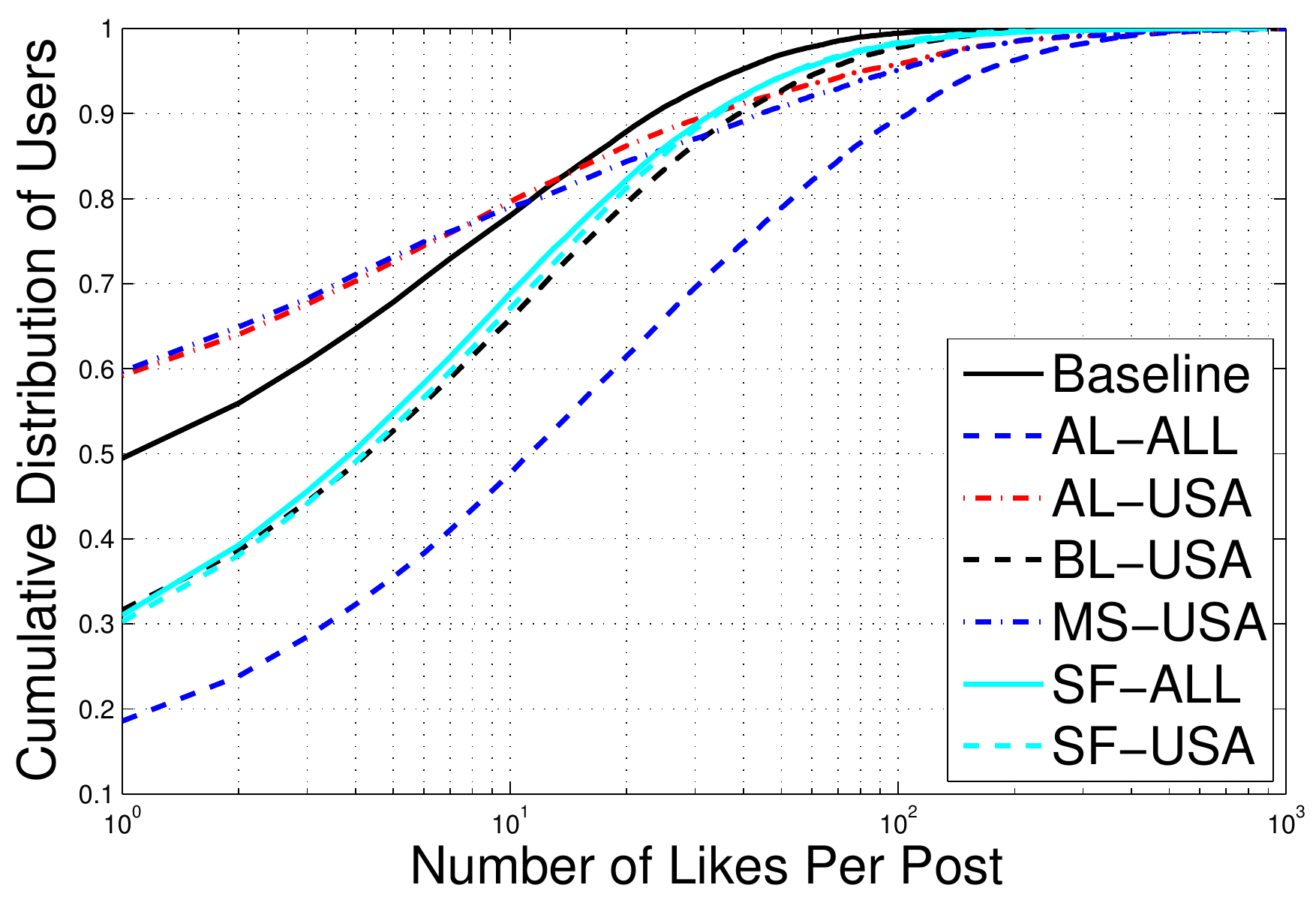}\label{fig:likes-post}}
\\
\subfigure[]{\includegraphics[width=0.35\columnwidth]{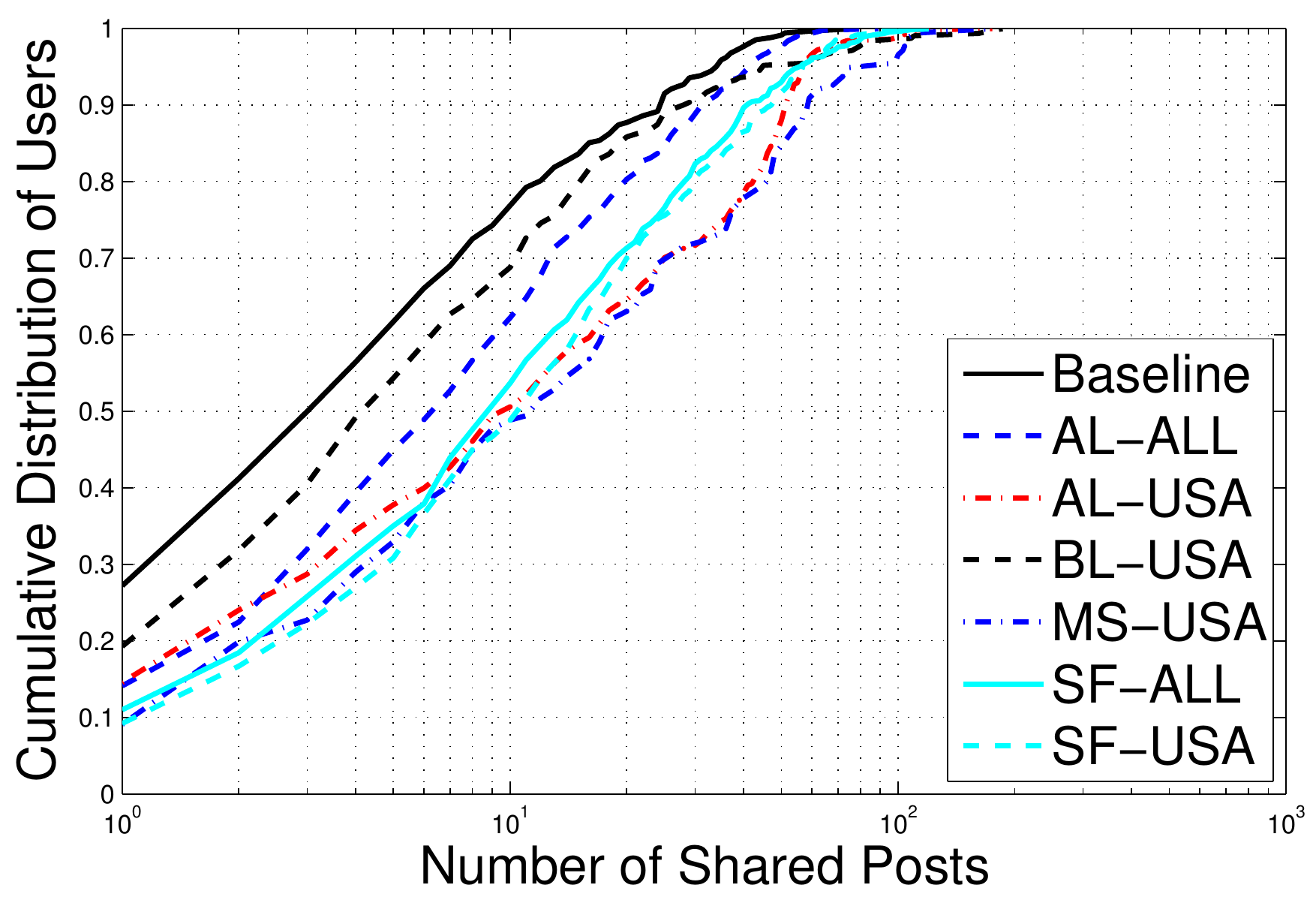}\label{fig:share}}
~\hspace{-0.2cm}
\subfigure[]{\includegraphics[width=0.35\columnwidth]{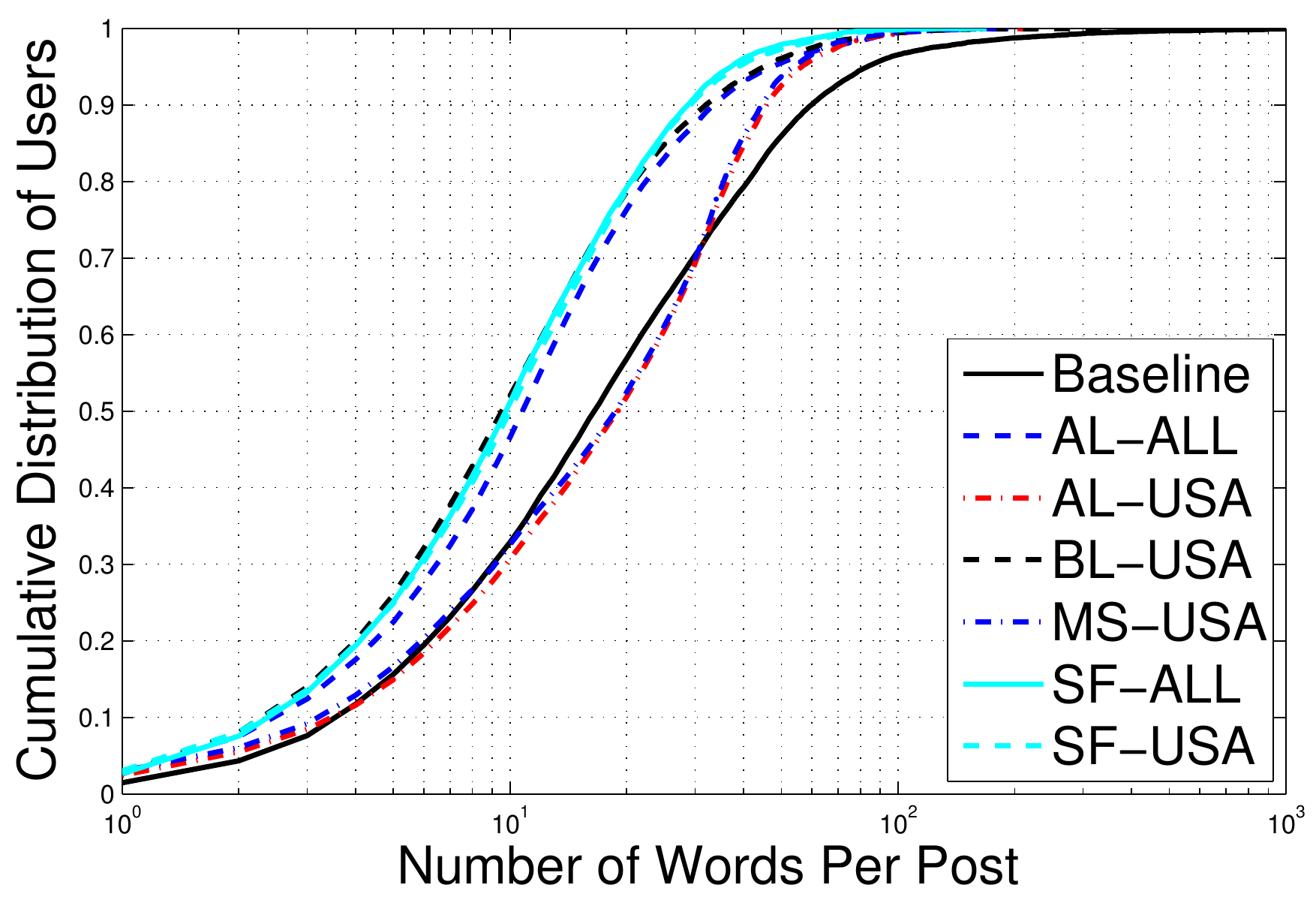}\label{fig:text}}
\vspace{-0.35cm}
\caption{Distribution of non-lexical features for like farm and baseline accounts}
\vspace{0.4cm}
\end{figure*}

\subsection{Analysis of Non-Lexical Features}
\label{sec:charNL}

\descr{Comments and Likes.} In Figure~\ref{fig:comperpost}, we plot the distributions of the number of comments a post attracts, revealing that users of AL-ALL like farm generate many more comments than the baseline users. We note that BL-USA is almost identical to the baseline users. Next, Figure~\ref{fig:likes-post} shows the number of likes associated with users' posts, highlighting that posts of like farm users attract much more likes than those of baseline users. Therefore, posts produced by the former gather more likes (and also have lower lexical richness as shown later on in Table~\ref{tab:lexicalfeatures}), which might actually indicate their attempt to mask suspicious activities.

\descr{Shared Content.} We next study the distributions of posts that are classified as ``shared activity," i.e.,  originally made by another user, or articles, images, or videos linked from an external URL (e.g., a blog or YouTube). Figure~\ref{fig:share} shows that baseline users generate more original posts, and share fewer posts or links, compared to farm users.

\descr{Words per Post.}
Figure~\ref{fig:text} plots the distributions of number of words that make up a text-based post, highlighting that posts of like farm users tend to have fewer words. Roughly half of the users in four of the like farms (AL-ALL, BL-USA, SF-ALL, and SF-USA) use 10 or less words in their posts, versus 17 words by baseline users.

\begin{figure*}[t]
	\centering
	\subfigure[AL]{
		\includegraphics[width=.35\textwidth]{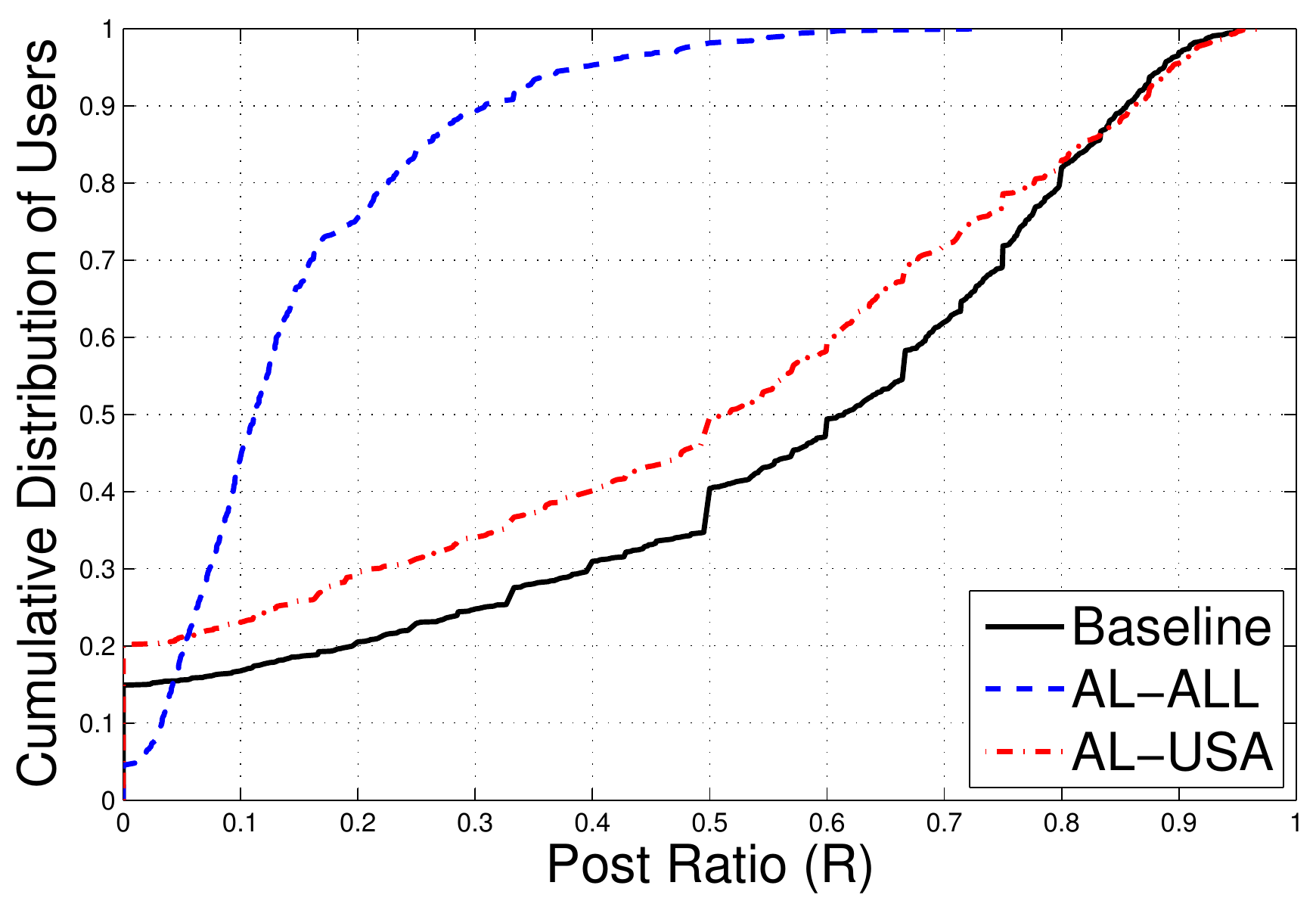}
		\label{fig:al_campaign}
	}\hspace{-0.3cm}
	~
	\subfigure[BL]{
		\includegraphics[width=.35\textwidth]{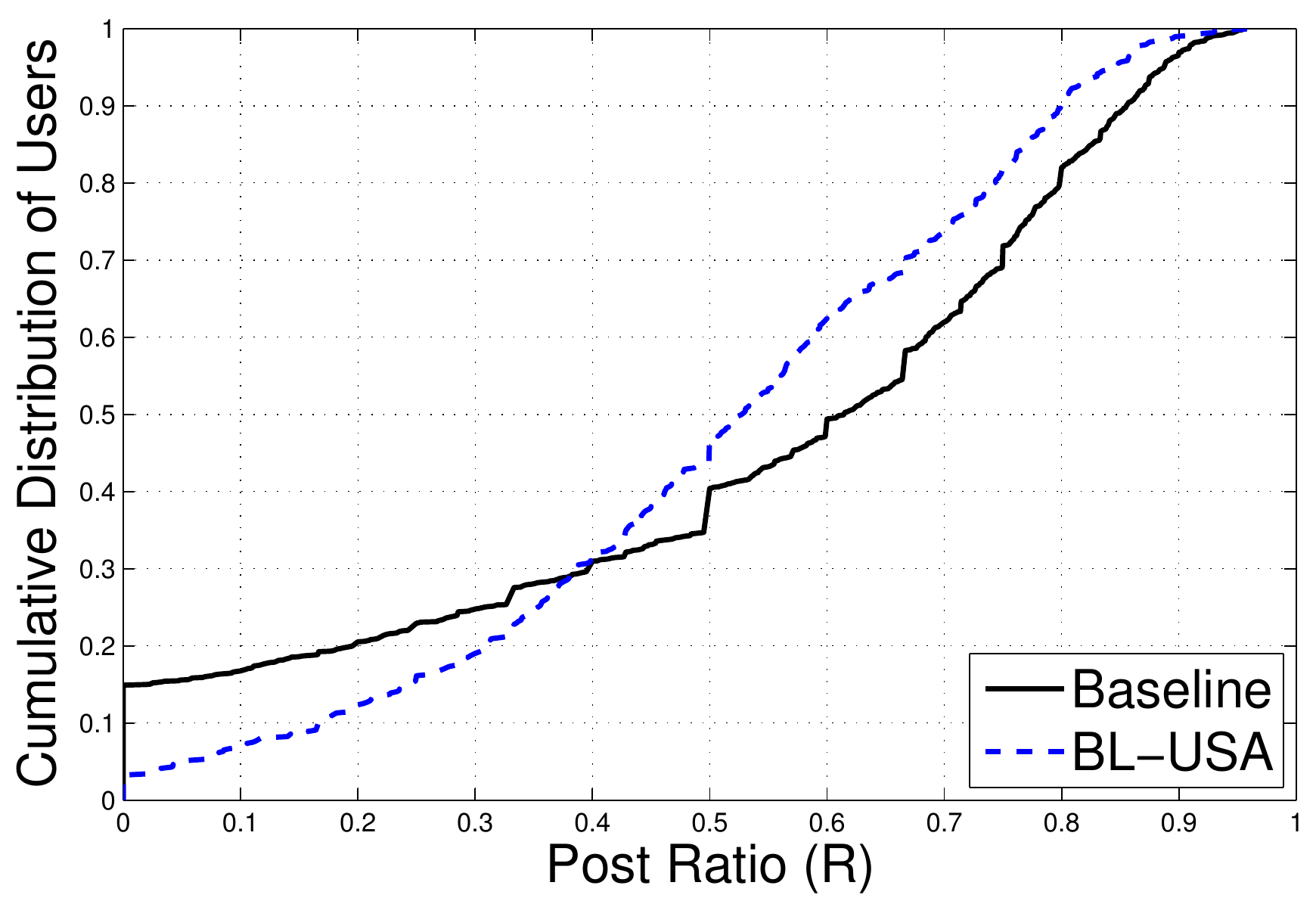}
		\label{fig:bl_campaign}
	}
	\\
	\subfigure[SF]{
		\includegraphics[width=.35\textwidth]{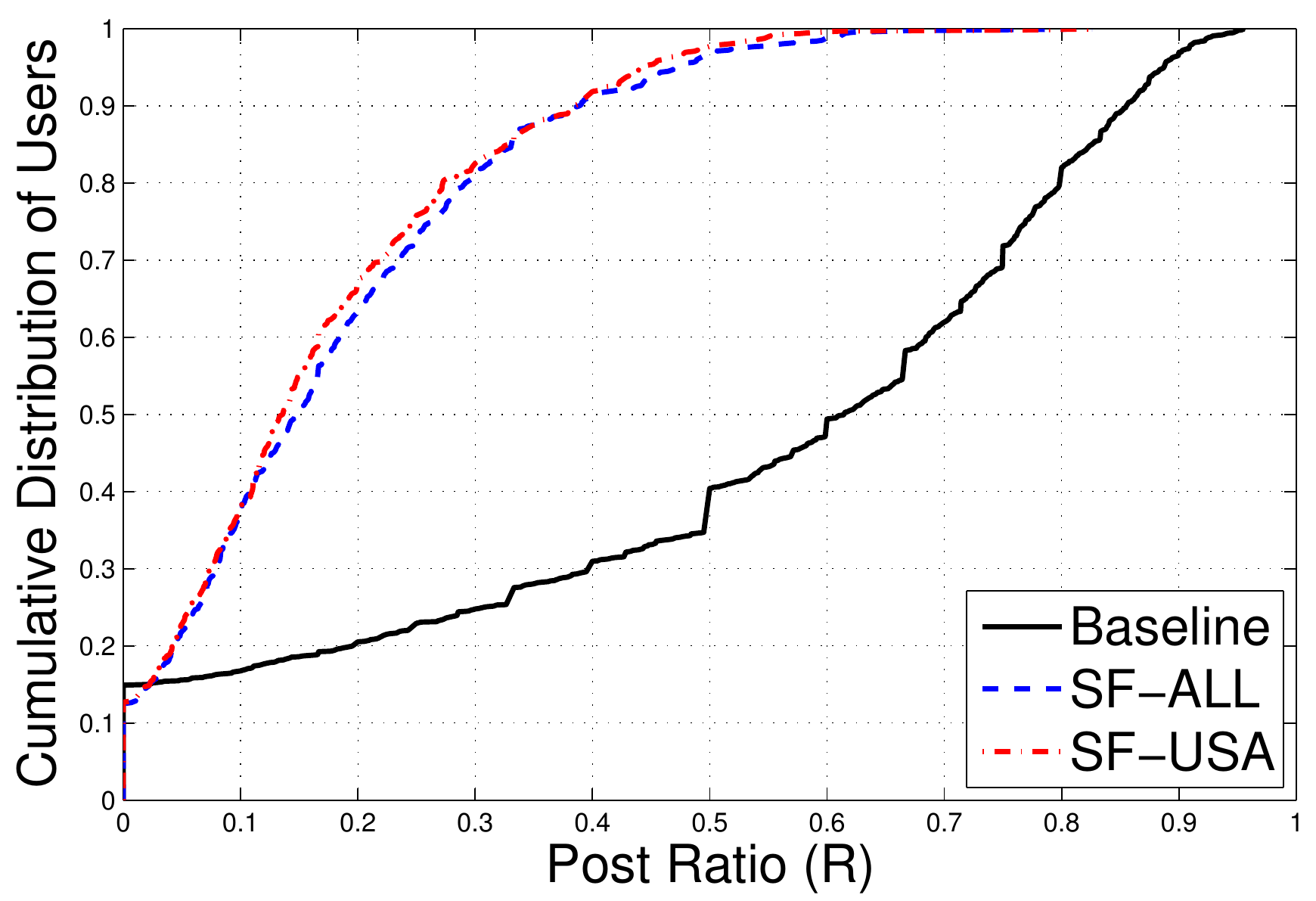}
		\label{fig:ms_us_n_sf}
	}\hspace{-0.3cm}
	~
	\subfigure[MS]{
		\includegraphics[width=.35\textwidth]{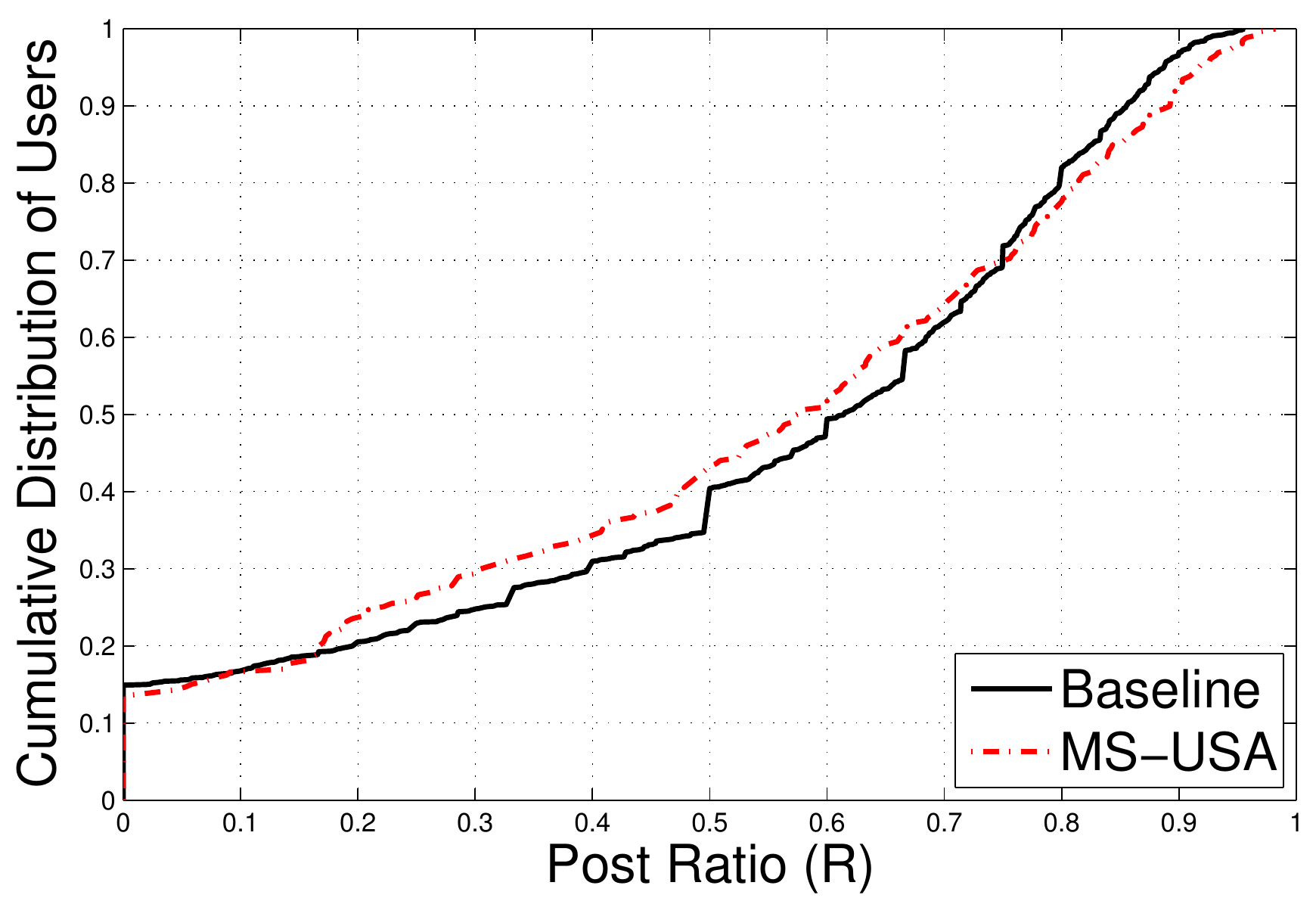}
		\label{fig:ms_us_n_sf}
	}

	\vspace{-0.2cm}
	\caption{Distributions of the ratio of English to non-English posts.}%
	\label{fig:lang_ratio}
\vspace{0.4cm}
\end{figure*}

\subsection{Analysis of Lexical Features}
\label{subsec:lexicalanalysis}
We now look at features that relate to the content of timeline posts. Although we only consider posts in English, similar lexical features could be extracted for other languages such as for Chinese~\cite{Zhang:2003:HCL:1119250.1119280}. The readers are referred to \cite{silberztein1997lexical} for more details about features used in this paper. We have also considered user timelines as the collection of posts and the corresponding comments on each post (i.e., all textual content) and build a corpus of words extracted from the timelines by applying the term frequency-inverse document frequency (TF-IDF) statistical tool~\cite{Salton:tfidf}. However, the overall performance of this ``bag-of-words'' approach was poor, which can be explained with the short nature of the posts. Indeed, \cite{Hogenboom:2015} has showed that the word frequency approach to analyze short text on social media and blogs does not perform well. Thus, in our work, we disregard simple TF-IDF based analysis of user timelines and identify other lexical features.

\descr{Language.} Next, we analyze the ratio of posts in English, i.e., for every post we filter out all non-English ones using a standard language detection library.\footnote{\url{https://python.org/pypi/langdetect}} For each user, we count the number of English-language posts and calculate its ratio with respect to the total number of posts. Figure~\ref{fig:lang_ratio} shows that the baseline users and like farm users in USA (i.e., MS-USA, BL-USA, and AL-USA) mostly post in English, while users of worldwide campaigns (MS-ALL, BL-ALL, AL-ALL) have significantly fewer posts in English. For example, the median ratio of English posts for AL-ALL campaign is around 10\% and that for SF-ALL around 15\%. We acknowledge that our analysis is limited to English-only content and may be statistically biased toward native English speakers i.e, non-USA users. While our analysis should be extended to other languages, we argue that English-based lexical analysis provides sufficient differences across different categories of users. Thus, developing algorithms for language detection and processing on non-English posts is out of the scope of this paper.

\descr{Readability.} We further analyze posts for grammatical and semantic correctness. We parse each post to extract the number of words, sentences, punctuation, non-letters (e.g., emoticons), and measure the lexical richness, as well as the Automated Readability Index (ARI)~\cite{smith:ari} and Flesch score \cite{flesch48readability}. Lexical richness, defined as the ratio of number of unique words to total number of words, reveals noticeable repetitions of distinct words, while the ARI, computed as 4.71 $\times$ average word length) + (0.5 $\times$ average sentence length) - 21.43, estimates the comprehensibility of a text corpus.
Table~\ref{tab:lexicalfeatures} shows a summary of the results. In comparison to like farm users, baseline users post text with higher lexical richness (70\% vs. 55\%), ARI (20 vs. 15), and Flesch score (55 vs. 48), thus suggesting that normal users use a richer vocabulary and that their posts have higher readability.

\begin{table*}[!t]
\begin{center}
\small
\tabcolsep=0.12cm
\begin{tabular}{lrrrrrrrrrrrr}
\toprule
&  	\bf Avg 	& \bf Avg & \bf Avg & {\bf Avg Sent} & {\bf Avg Word}  &  &  & {\bf Flesch}\\[-0.5ex]
{\bf Campaign} &  {\bf Chars}	&  {\bf Words}	& {\bf Sents} &  {\bf Length}  & {\bf Length} & {\bf Richness} & {\bf ARI} & {\bf Score}\\
\midrule
Baseline	&	4,477	&	780	&	67	&	6.9	&	17.6	&	\textbf{0.70}	&	\textbf{20.2}	&	55.1	\\
\midrule

BL-USA	&	7,356	&	1,330	&	63	&	5.7	&	22.8	&	0.58	&	16.9	&	51.5	\\
AL-ALL	&	2,835	&	464	&	32	&	6.2	&	13.9	&	0.59	&	14.8	&	43.6	\\
AL-USA	&	2,475	&	394	&	33	&	6.2	&	12.7	&	0.49	&	14.1	&	54.0	\\
SF-ALL	&	1,438	&	227	&	19	&	6.3	&	11.7	&	0.58	&	14.1	&	45.2	\\
SF-USA	&	1,637	&	259	&	22	&	6.3	&	12.0	&	0.55	&	14.4	&	45.6	\\
MS-USA	&	6,227	&	1,047	&	66	&	6.1	&	17.8	&	0.53	&	16.2	&	50.1	\\
\bottomrule
\end{tabular}
\precaption\precaption		
\caption{Lexical analysis of timeline posts.}
\label{tab:lexicalfeatures}		
\vspace{-0.25cm}
\end{center}
\end{table*}

\subsection{Summary \& Takeaways}
Our analysis of user timelines highlights several differences in both lexical and non-lexical features of normal and like farm users. In particular, we find that posts made by like farm accounts have 43\% fewer words, a more limited vocabulary, and lower readability than normal users' posts. Moreover, %
like farm users generate significantly more comments and likes and a large fraction of their posts consists of non-original and often redundant ``shared activity''. 

In the next section, we will use these timelines features to automatically detect like farm users using a machine learning classifier.

\section{Timeline-based Detection of Like Farms}
\label{sec:detection}
Aiming to automatically distinguish like farm users from normal (baseline) users, we use a supervised two-class SVM classifier~\cite{Muller01anintroduction}, implemented using \emph{scikit-learn}~\cite{sklearn_api} (an open source machine learning library for Python). We later compare this classifier with other well-known supervised supervisor such as Decision Tree~\cite{dtree}, AdaBoost~\cite{adaboost}, kNN~\cite{knn}, Random Forest~\cite{Breiman:rf}, and Na\"ive Bayes~\cite{zhang2004optimality} and confirm that the two-class SVM is the most effective in detecting like farms users.

We extract four non-lexical features and twelve distinct lexical features from the timelines of baseline and like farm users, as explained in Section~\ref{sec:characterizing}. The non-lexical features are the average number of words, comments, likes per post, and re-shares. The lexical features include: the number of characters, words, and sentences; the average word length, sentence length, and number of upper case letters; the average percentage of punctuation, numbers, and non-letter characters; richness, ARI, and Flesch Score.

We form two classes by labeling like farm and baseline users' lexical and non-lexical features as positives and negatives, respectively. We use 80\% and 20\% of the features to build the training and testing sets, respectively. Appropriate values for parameters $\gamma$ (\emph{radial basis function kernel} parameter~\cite{Scholkopf:2001:ESH:1119748.1119749}) and $\upsilon$ (SVM parameter) are set empirically by performing a greedy grid search on ranges $2^{-10} \leq \gamma \leq 2^{0}$ and $2^{-10} \leq \upsilon \leq 2^{0}$, respectively, on each training group.

\begin{table*}[t]
\tabcolsep=0.11cm
  \begin{center}
	\small
    \begin{tabular}{crrrrrrrrrrr}
      \toprule
			\bf Campaign	&	\bf Total	&	\bf  Training	&	\bf Testing 	&	\bf TP	&	\bf FP	&	\bf TN	&	\bf FN	&	\bf Precision	&	\bf Recall	&	\bf Accuracy	 &	\bf F1-\\[-0.5ex]
& \bf Users & \bf Set & \bf Set & & & & & & & & \bf Score \\
\midrule

BL-USA	&	583	&	466	&	117	&	37	&	12	&	270	&	80	&	76\%	&	32\%	&	77\%	&	45\%	\\
AL-ALL	&	707	&	566	&	141	&	132	&	5	&	278	&	9	&	96\%	&	94\%	&	97\%	&	95\%	\\
AL-USA	&	827	&	662	&	164	&	113	&	4	&	278	&	51	&	97\%	&	69\%	&	88\%	&	81\%	\\
SF-ALL	&	870	&	696	&	174	&	139	&	9	&	273	&	35	&	94\%	&	80\%	&	90\%	&	86\%	\\
SF-USA	&	653	&	522	&	131	&	110	&	5	&	277	&	21	&	96\%	&	84\%	&	94\%	&	90\%	\\
MS-USA	&	259	&	207	&	52	&	39	&	2	&	280	&	13	&	95\%	&	75\%	&	96\%	&	84\%	\\
\bottomrule
    \end{tabular}
 
	\precaption
    \caption{Effectiveness of non-lexical features (+SVM) in detecting like farm users.}
    \label{tab:svm_on_non_lexical_features}
	\vspace{-0.6cm}
      \end{center}
\end{table*}

\descr{Non-Lexical Features.}
Table~\ref{tab:svm_on_non_lexical_features} reports on the accuracy of our classifier with non-lexical features, i.e., users interactions with posts as described in Section~\ref{sec:charNL}. Note that for each campaign, we train the classifier with 80\% of the non-lexical features from baseline and campaign training sets derived from the campaign users timelines. The poor classification performance for the stealthiest like farm (BL-USA) suggests that non-lexical features alone are not sufficient to accurately detect like farm users.

\begin{table*}[t]
\tabcolsep=0.11cm
  \begin{center}
   \small
    \begin{tabular}{crrrrrrrrrrr}
      \toprule
			\bf Campaign	&	\bf Total	&	\bf  Training	&	\bf Testing 	&	\bf TP	&	\bf FP	&	\bf TN	&	\bf FN	&	\bf Precision	&	\bf Recall	&	\bf Accuracy	 &	\bf F1-	\\[-0.5ex]
& \bf Users & \bf Set & \bf Set & & & & & & & & \bf Score \\
\midrule
BL-USA	&	564	&	451	&	113	&	113	&	0	&	240	&	0	&	100\%	&	100\%	&	100\%	&	100\%	\\
AL-ALL	&	675	&	540	&	135	&	129	&	2	&	238	&	6	&	98\%	&	96\%	&	98\%	&	97\%	\\
AL-USA	&	570	&	456	&	114	&	113	&	0	&	240	&	1	&	100\%	&	99\%	&	99\%	&	99\%	\\
SF-ALL	&	761	&	609	&	152	&	150	&	1	&	239	&	2	&	99\%	&	99\%	&	99\%	&	99\%	\\
SF-USA	&	570	&	456	&	114	&	99	&	2	&	238	&	15	&	98\%	&	87\%	&	95\%	&	92\%	\\
MS-USA	&	224	&	179	&	45	&	45	&	0	&	240	&	0	&	100\%	&	100\%	&	100\%	&	100\%	\\
\bottomrule
    \end{tabular}
  
	\precaption
      \caption{Effectiveness of lexical features (+SVM) in detecting like farm users.}
      \label{tab:svm_on_lexical_features}
    \vspace{-0.3cm}
  \end{center}
\end{table*}

\descr{Lexical Features.}
Next, we evaluate the accuracy of our classifier with lexical features, reported in Table~\ref{tab:svm_on_lexical_features}. We filter out all users with no English-language posts (i.e, with R=0, see Figure~\ref{fig:lang_ratio}). Again, we train the classifier with 80\% lexical features from baseline and like farm training sets. We observe that our classifier achieves very high precision and recall for MS-USA, BL-USA, and AL-USA. Although the accuracy decreases by approximately 8\% for SF-USA, the overall performance suggests that lexical features are useful in automatically detecting like farm users.

\begin{table*}[!t]
\tabcolsep=0.11cm
\small
  \begin{center}
    \begin{tabular}{crrrrrrrrrrr}
      \toprule
			\bf Campaign	&	\bf Total	&	\bf  Training	&	\bf Testing 	&	\bf TP	&	\bf FP	&	\bf TN	&	\bf FN	&	\bf Precision	&	\bf Recall	&	\bf Accuracy	 &	\bf F1-	\\[-0.5ex]
& \bf Users & \bf Set & \bf Set & & & & & & & & \bf Score \\
\midrule
BL-USA	&	583	&	466	&	117	&	113	&	1	&	281	&	4	&	99\%	&	97\%	&	99\%	&	98\%	\\
AL-ALL	&	707	&	566	&	141	&	137	&	1	&	281	&	4	&	99\%	&	97\%	&	99\%	&	98\%	\\
AL-USA	&	827	&	662	&	164	&	157	&	1	&	281	&	7	&	99\%	&	96\%	&	98\%	&	97\%	\\
SF-ALL	&	870	&	696	&	174	&	163	&	2	&	280	&	11	&	99\%	&	94\%	&	97\%	&	96\%	\\
SF-USA	&	653	&	522	&	131	&	122	&	1	&	281	&	9	&	99\%	&	93\%	&	98\%	&	96\%	\\
MS-USA	&	259	&	207	&	52	&	50	&	0	&	282	&	2	&	100\%	&	96\%	&	99\%	&	98\%	\\
\bottomrule
    \end{tabular}
	\precaption
    \caption{Effectiveness of both lexical and non-lexical features (+SVM) in detecting like farm users.}
    \label{tab:svm_on_all_features}
    \vspace{-0.6cm}
      \end{center}
\end{table*}

\descr{Combining Lexical and Non-Lexical Features.}
Approximately 3\% to 22\% of like farm users and 14\% of baseline users do not have English language posts and are not considered in the lexical features based classification. To include them %
in the classification, for each like farm and baseline, we set their lexical features to zeros and aggregate the lexical features with non-lexical features, and evaluate our classifier with the same classification methodology as detailed above. Results are summarized in Table~\ref{tab:svm_on_all_features}, which shows high accuracy for all like farms (F1-Score $\geq$ 96\%), thus confirming the effectiveness of our timeline-based features in detecting like farm users.

\begin{table*}[!t]
\tabcolsep=0.11cm
\small
  \begin{center}
    \begin{tabular}{crrrrrr}
      \toprule
{\bf Campaign}	 & \bf SVM& \bf Decision Tree	& \bf AdaBoost & \bf	kNN	& \bf Random Forest & \bf Na\"ive Bayes\\
\midrule
BL-USA	&	98\%	&	96\%	&	96\%	&	91\%	&	88\%	&	53\%	\\
AL-ALL	&	98\%	&	84\%	&	95\%	&	86\%	&	84\%	&	75\%	\\
AL-USA	&	97\%	&	88\%	&	90\%	&	91\%	&	86\%	&	81\%	\\
SF-ALL	&	96\%	&	90\%	&	94\%	&	89\%	&	87\%	&	67\%	\\
SF-USA	&	96\%	&	83\%	&	92\%	&	79\%	&	78\%	&	61\%	\\
MS-USA	&	98\%	&	90\%	&	89\%	&	89\%	&	87\%	&	74\%	\\
\bottomrule
    \end{tabular}

	\precaption
    \caption{F1-Score with different classification methods, using both lexical and non-lexical features, in detecting like farm users.
}
    \label{tab:classifiers_on_all_features}
    \vspace{-0.2cm}
      \end{center}
\end{table*}

\descr{Comparison With Other Machine Learning Classifiers.}
In order to generalize our approach, we have also used other machine learning classification algorithms, i.e., Decision Tree, AdaBoost, kNN, Random Forest, and Na\"ive Bayes. The training and testing of all these classifiers follow the same set-up as the SVM approach. We again use 80\% and 20\% of the combined lexical and non-lexical features to build the training and testing sets, respectively. We summarize the performance of the classifiers in Table~\ref{tab:classifiers_on_all_features}.  Our results show that the SVM classifier achieves the highest F1-Scores across the board. Due to overfitting on our dataset, Random Forest and Na\"ive Bayes show poor results and require mechanism such as pruning, detailed analysis of parameters, as well as selection of the optimal set of prominent features to improve classification performance~\cite{Breiman:rf}~\cite{kohavi1995feature}.

\begin{figure*}[!t]
	\centering
\subfigure[]{\includegraphics[width=0.38\columnwidth,height=5cm]{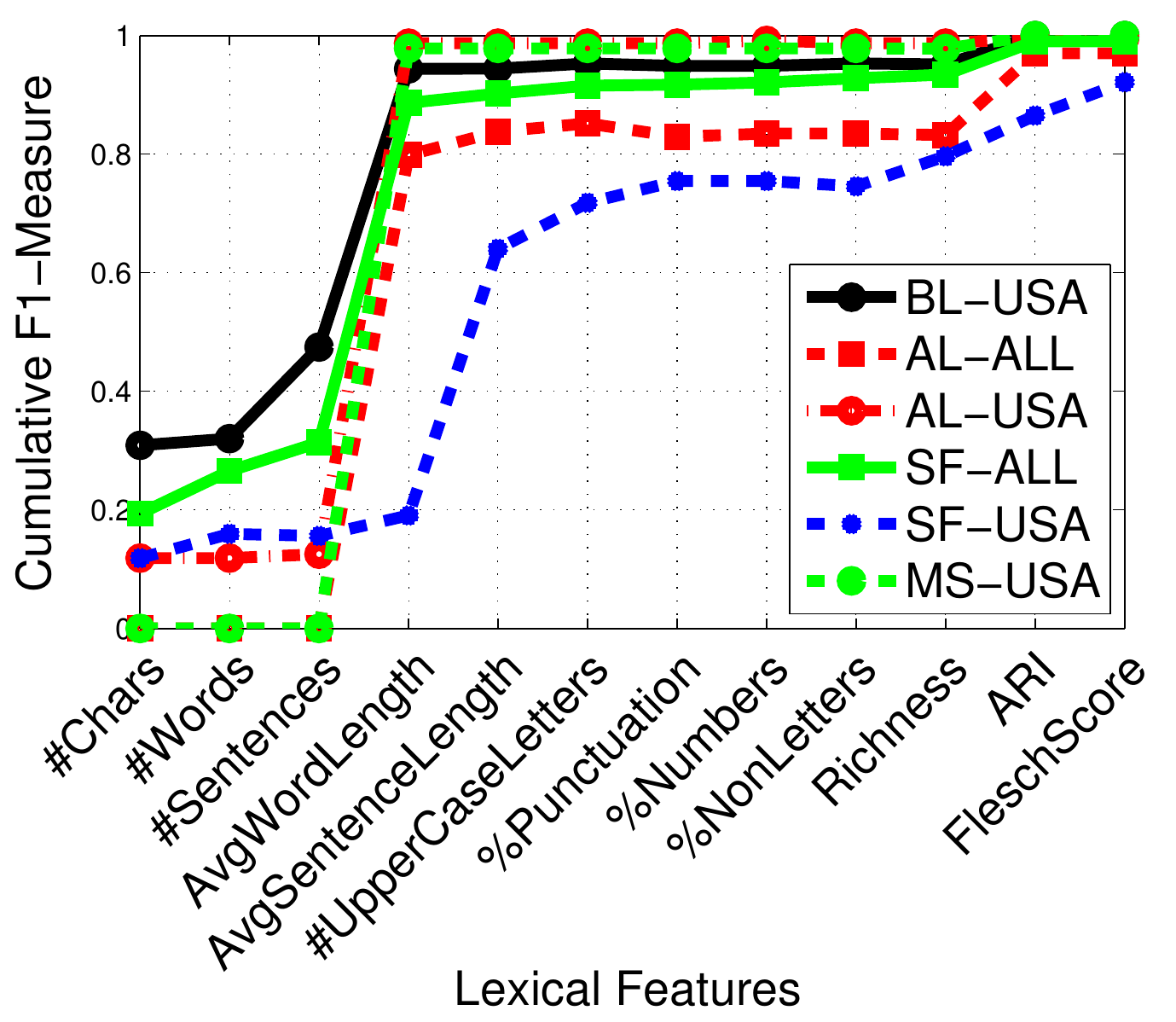}\label{fig:lexical_fa}}
~\hspace{-0.4cm}
\subfigure[]{\includegraphics[width=0.40\columnwidth,height=5cm]{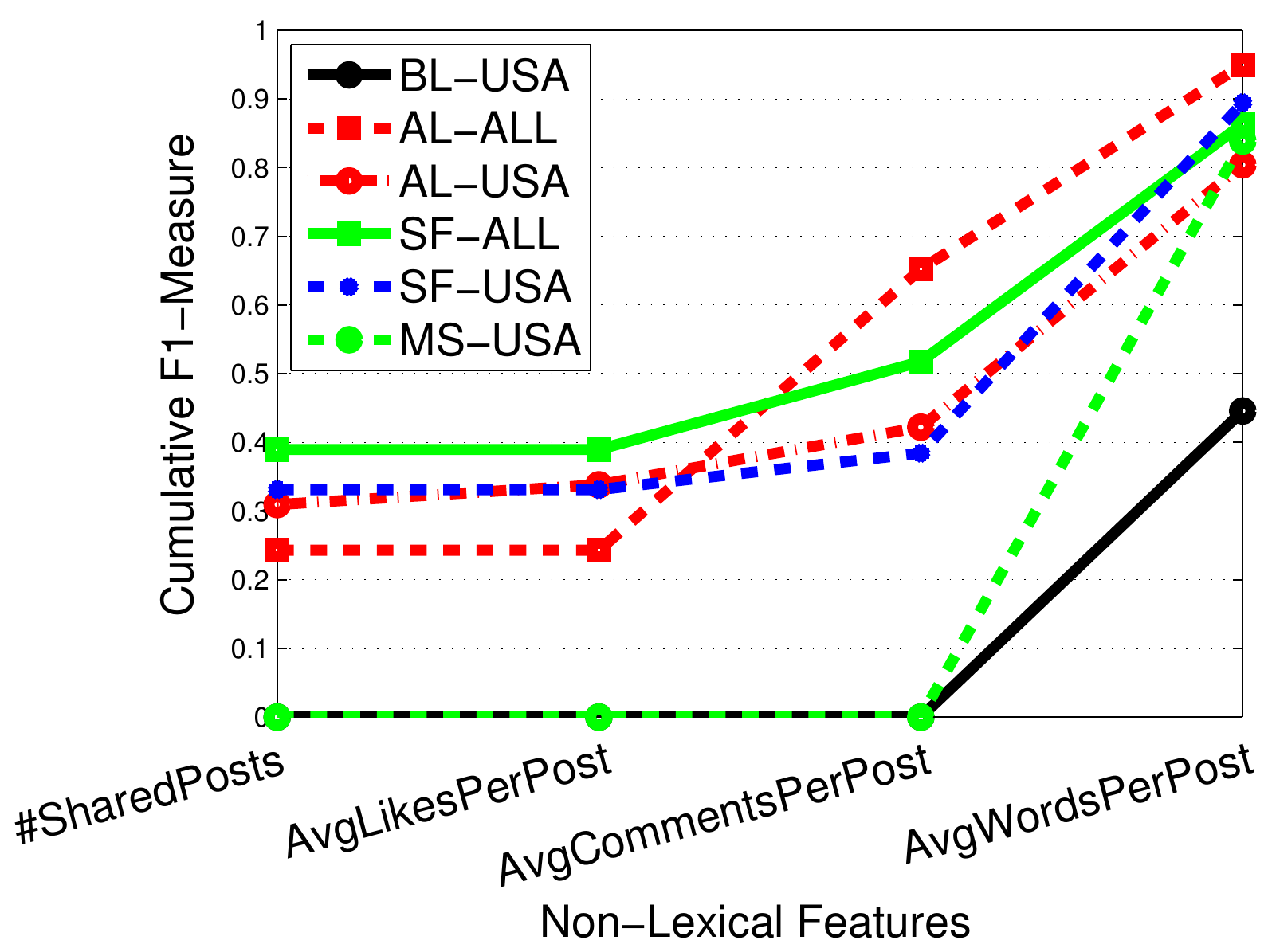}\label{fig:nonlexical_fa}}\\
\subfigure[]{\includegraphics[width=0.45\columnwidth,height=6cm]{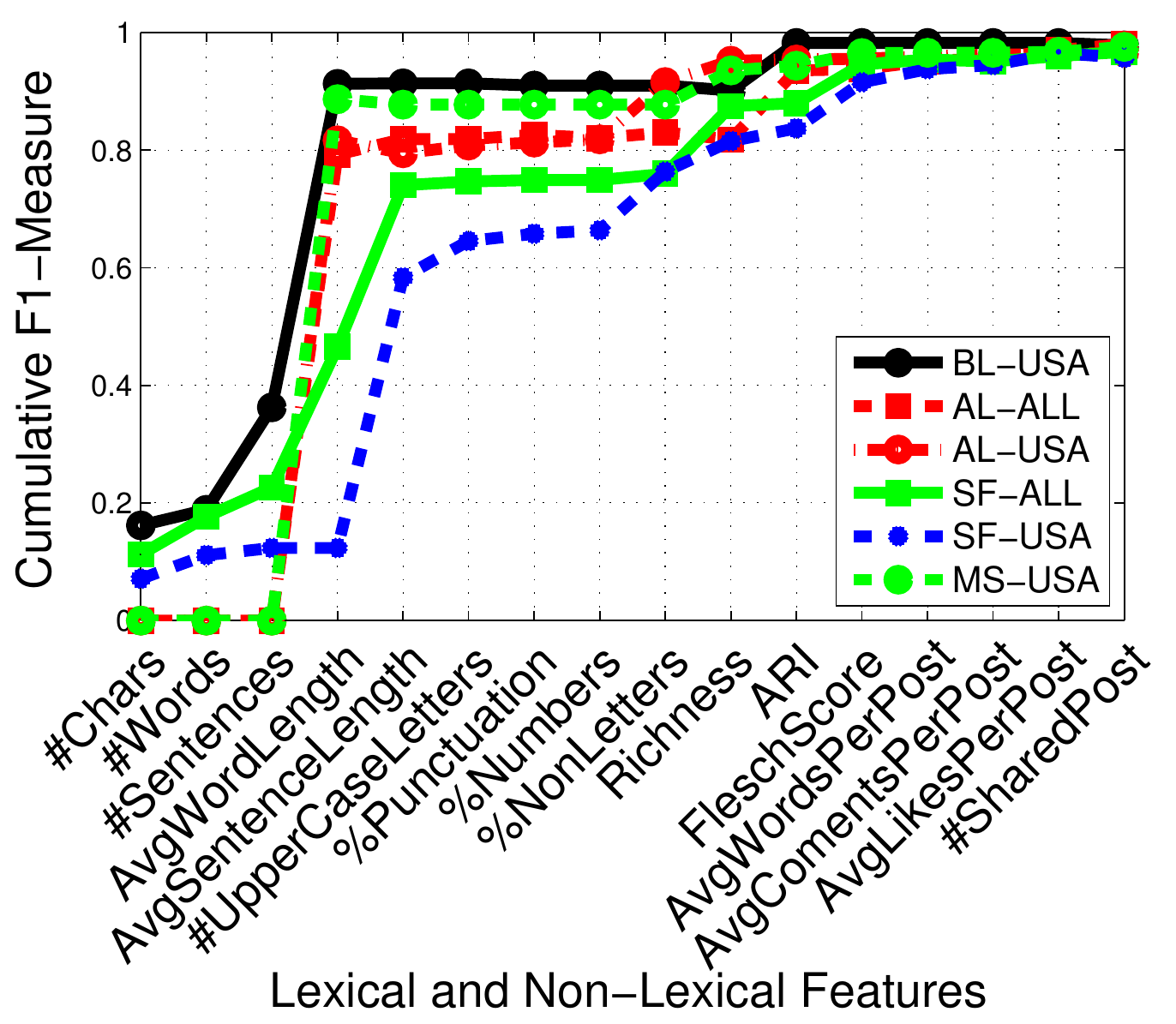}\label{fig:combined_fa}}
\vspace{-0.25cm}
\caption{Cumulative F1-Score for all lexical and non-lexical features measured. The X-axis shows the incremental inclusion of features in both training and testing of SVM. Details of the classification performance for all features are listed in Table~\ref{tab:svm_on_all_features}.}
\label{fig:factoranalysis}
\vspace{0.5cm}
\end{figure*}

\descr{Analysis.}
We now analyze in more details the classification performance (in terms of F1-Score) to identify the most distinctive features. Specifically, we incrementally add lexical and non-lexical features to train and test our classifier for all campaigns. We observe that the average word length (cf. Figure~\ref{fig:lexical_fa}) and average number of words per post (cf. Figure~\ref{fig:nonlexical_fa}) provide the most improvement in the F1-Score for all campaigns. This finding suggests that like farm users use shorter words and fewer number of words in their timeline posts as compared to baseline users. While these features provide the largest improvement in detecting a like farm account, an attempt to circumvent detection by increasing the word length or number of words per post will also effect the ARI, Flesch score, and richness. That is, increasing word length and number of words on posts in a way that is not readable nor understandable, will not improve the overall outlook of the account to appear real. Therefore, combining several features increases the workload required to appear real on like farm accounts. %
The overall classification accuracy with both lexical and non-lexical features is reported in Figure~\ref{fig:combined_fa}.

\descr{Remarks.}
Our results demonstrate that it is possible to accurately detect like farm users from both sophisticated and na\"ive farms by incorporating additional account information -- specifically, timeline activities. We also argue that the use of a variety of lexical and non-lexical features will make it difficult for like farm operators to circumvent detection. Like farms typically rely on pre-defined lists of comments, resulting in word repetition and lower lexical richness. As a result, we argue that, should our proposed techniques be deployed by Facebook, it will be challenging, as well as costly, for fraudsters to modify their behavior and evade detection, since this would require instructing automated scripts and/or cheap human labor to match the diversity and richness of real users' timeline posts.

\section{Related Work}
\label{sec: related work}

Prior work has focused quite extensively on the analysis and the detection of fake accounts in online social networks \cite{boshmaf15integro,cao12fakeosn,gao10socialspamcampaigns,yang12spammersocialnetwork,yang11socialnetworksybils}.
By contrast, we focus on detecting accounts that are employed by like farms to boost the number of Facebook page likes, whether they are operated by a bot or a human.

Our classifier is trained using accounts obtained from honeypots, somewhat similar to previous work 
in the context of spam in MySpace and Twitter~\cite{lee10socialspamhoneypots,stringhini10spammerssocialnetworks}. %
Our work uses accounts attracted by Facebook pages actively engaging like farms %
and, unlike~\cite{lee10socialspamhoneypots,stringhini10spammerssocialnetworks}, 
leverages timeline-based features for the detection.
Wang et al.~\cite{wang14adversarialdetection} study the human involvement in Weibo's reputation manipulation services,
showing that simple evasion attacks (e.g., workers modifying their behavior) as well as poisoning attacks (e.g., administrators tampering with the training set) can severely affect the effectiveness of machine learning algorithms to detect malicious crowd-sourcing workers. 
Partially informed by their work, we do not only cluster like activity performed by users but also build on
lexical and non-lexical features.

Other studies have analyzed services that sell {\em Twitter followers}~\cite{stringhini12twitterfollowermarketWOSN}, fake and compromised Twitter accounts~\cite{thomas13traffickingfraudtwitteraccounts}, as well as {\em crowdturfing} in social networks~\cite{song2015crowdtarget}.
Specific to Facebook fraud is CopyCatch~\cite{beutel2013copycatch}, 
a technique currently deployed by Facebook to detect fraudulent accounts by identifying groups of connected users liking a set of pages within a short time frame. 
SynchroTrap~\cite{cao14synchrotrap} extends CopyCatch by clustering accounts that perform similar, possibly malicious, synchronized actions, using tunable parameters such as time-window and similarity thresholds in order to improve detection accuracy. 
However, as highlighted in our prior work~\cite{decristofaro14facebooklikefarms}, while some farms seem to be operated by bots (producing large bursts of likes and having limited numbers of friends) that do not really trying to hide their activities, others, {\em stealthier} farms exhibit behavior that may be challenging to detect with tools like CopyCatch and SynchroTrap.
In fact, our evaluation of graph co-clustering techniques shows that these farms successfully evade detection by avoiding lockstep behavior and liking sets of seemingly random pages. 
As a result, we decide to use timeline features, relying on both lexical and non-lexical features to build a classifier  detecting stealthy like farm users with high accuracy.
Our work can complement other methods used in prior work to detect fake and compromised accounts, such as using unsupervised anomaly detection techniques~\cite{viswanath14tanomaloussocialnetwork}, temporal features~\cite{jiang14catchsyn,jiang14strangebehaviorosocial}, or IP addresses~\cite{stringhini2015evilcohort}.

Finally, we stress that our prior work~\cite{decristofaro14facebooklikefarms} only presents an exploratory measurement study of like farms, based on the characteristics of the accounts that liked a few honeypot pages. Specifically,~\cite{decristofaro14facebooklikefarms} analyzes the geographic and demographic  distribution of garnered likes, the temporal patterns observed for each campaign, as well as the social graph induced by the likers. %
Whereas, in this paper, we take a significant step further: although we re-use the honeypot campaigns to build a corpus of like farm users, (i) we  demonstrate that temporal and social graph analysis can only be used to detect naive farms, and
(ii) we introduce a timeline-based classifier that achieves a remarkably high degree of accuracy.

\section{Conclusion}
\label{sec:conclusion}
The detection of fraudulent accounts in online social networks is crucial to maintain confidence among legitimate users and investors. In this paper, we focused on detecting accounts used by Facebook like farms, i.e., paid services artificially boosting the number of likes on a given Facebook page. We crawled liking patterns and timeline activities from like farms accounts and from a baseline of normal users. We evaluated the effectiveness of existing graph based fraud detection algorithms, such as CopyCatch~\cite{beutel2013copycatch} and SynchroTrap~\cite{cao14synchrotrap}, and demonstrated that sophisticated like farms can successfully evade detection.

Aiming to address this problem, we set to incorporate additional profile information from accounts' timelines to train machine learning classifiers geared to distinguish between like farm users from normal ones. We first experimented with term frequency-inverse document frequency (TF-IDF) but achieve relatively poor performance. We then turned to lexical and non-lexical features from user timelines. We found that posts made by like farm accounts have 43\% fewer words, a more limited vocabulary, and lower readability than normal users' posts. Moreover, like farm posts generated significantly more comments and likes, and a large fraction of their posts consists of non original and often redundant ``shared activity'' (i.e., repeatedly sharing posts made by other users, articles, videos, and external URLs). By leveraging both lexical and non-lexical features, we experimented with several machine learning classifiers, with the best of our classifiers (SVM) achieving as high as 100\% precision and 97\% of recall, and at least 99\% and 93\% respectively across all campaigns -- significantly higher than graph co-clustering techniques.

In theory, fraudsters could try to modify their behavior in order to evade our proposed timeline-based detection. However, like farms either heavily automate mechanisms or rely on manual input of cheap human labor. Therefore, since non-lexical features are extracted from users' interactions with timeline posts, imitating normal users' behaviors will likely incur an remarkably higher cost. Even higher would be the cost to interfere with lexical features, since this would entail modifying or imitating normal users' writing style.

\end{document}